\documentclass[pdflatex,sn-basic, Numbered]{sn-jnl}

\usepackage{graphicx}%
\usepackage{multirow}%
\usepackage{amsmath,amssymb,amsfonts}%
\usepackage{amsthm}%
\usepackage{mathrsfs}%
\usepackage[title]{appendix}%
\usepackage{xcolor}%
\usepackage{textcomp}%
\usepackage{manyfoot}%
\usepackage{booktabs}%
\usepackage{algorithm}%
\usepackage{algorithmicx}%
\usepackage{algpseudocode}%
\usepackage{listings}%

\usepackage{hyperref}
\usepackage{placeins}
\usepackage{multirow}
\usepackage{makecell}
\usepackage{booktabs}
\usepackage{color,soul}

\begin{document}

\title[Article Title]{Understanding Trends, Patterns, and Dynamics in Global Company Acquisitions: A Network Perspective}


\author*[1]{\fnm{Ghazal} \sur{Kalhor}}\email{kalhor.ghazal@ut.ac.ir}

\author[2]{\fnm{Behnam} \sur{Bahrak}}\email{b.bahrak@teias.institute}

\affil*[1]{\orgdiv{School of Electrical and Computer Engineering, College of Engineering}, \orgname{University of Tehran}, \orgaddress{\city{Tehran}, \country{Iran}}}

\affil[2]{\orgname{Tehran Institute for Advanced Studies}, \orgaddress{\city{Tehran}, \country{Iran}}}


\abstract{Studying acquisitions offers invaluable insights into startup trends, aiding informed investment decisions for businesses. However, the scarcity of studies in this domain prompts our focus on shedding light in this area. Employing Crunchbase data, our study delves into the global network of company acquisitions using diverse network analysis techniques. Our findings unveil an acquisition network characterized by a primarily sparse structure comprising localized dense connections. We reveal a prevalent tendency among organizations to acquire companies within their own country and industry, as well as those within the same age bracket. Furthermore, we show that the country, region, city, and category of the companies can affect the formation of acquisition relationships between them. Our temporal analysis indicates a growth in the number of weakly connected components of the network over time, accompanied by a trend toward a sparser network. Through centrality metrics computation in the cross-city acquisition network, we identify New York, London, and San Francisco as pivotal and central hubs in the global economic landscape. Finally, we show that the United States, United Kingdom, and Germany are predominant countries in international acquisitions. The insights from our research assist policymakers in crafting better regulations to foster global economic growth, and aid businesses in deciding which startups to acquire and which markets to target for expansion.}

\keywords{International acquisitions, Network analysis, Startup trends, Crunchbase data, Economic hubs}



\maketitle

\section{Introduction}\label{sec1}

Today, a multitude of startups are emerging at a faster pace compared to previous times \cite{stayton2019seed}. This surge can be attributed, in part, to advancements in technology, leading to the rapid establishment of a great number of IT-based startups \cite{santisteban2017systematic}. The abundance of startups presents venture capitalists and investors with a wide array of options for investment or acquisition. Therefore, it is crucial for these organizations to conduct thorough research before making such decisions, aiming to enhance their prospects of success \cite{sreejesh2014business, raut2020past}. On the other hand, being acquired by a larger and financially stronger company is considered a measure of success and the ultimate goal for startups, making it a critical step for the founders of these young firms \cite{ferrati2021startup, cotei2018m}. As a result, offering a more profound understanding of global-level acquisitions holds great importance for businesses in various sectors and geographic regions, encompassing both large companies and startups at different stages of evolution.

Crunchbase\footnote{\url{https://www.crunchbase.com}}, PitchBook\footnote{\url{https://pitchbook.com}}, and VentureSource\footnote{\url{https://www.venture-source.com}} are among the most popular platforms providing extensive datasets, collecting information about global firms and their transactions, making them valuable data sources for evaluating acquisitions \cite{retterath2020benchmarking}. However, Crunchbase stands out as the most complete and comprehensive database among them, covering the greatest number of entries and features \cite{farber2019linked}. The Crunchbase team gathers data through various channels, including monthly portfolio updates from a global network of large investment firms, as well as contributions from executives, entrepreneurs, and investors who actively update company profile pages on Crunchbase. Additionally, they utilize artificial intelligence, machine learning algorithms, and data analysts to verify the accuracy of the data both automatically and manually\footnote{\url{https://support.crunchbase.com/hc/en-us/articles/360009616013-Where-does-Crunchbase-get-their-data}}. Crunchbase data provides information about companies and their transactions from all around the world, covering a wide range of industries from technology and life sciences to finance and consumer goods. Furthermore, this platform offers detailed information regarding acquisitions, including type, date, and price, making it a useful data source for analyzing various facets of these transactions. Therefore, we leverage this insightful data for our research to provide a holistic view of company acquisitions.

Social network analysis (SNA) has many applications, ranging from assessing robustness in complex networks to examining signal propagation within them \cite{artime2024robustness, ji2023signal}. In this study, we use SNA to construct and investigate the acquisition network of companies, where nodes represent the companies involved in acquisition transactions, and directed edges represent the connections from acquirer companies to acquirees. Then we explore the projections of this network onto acquiring and acquired firms, cities, and countries using Crunchbase data. We employ a network science approach in our analyses and interpret various structural features of these networks to gain a deeper understanding of dynamics and trends in acquisitions over time. 

Throughout this research work, our goal is to address the following questions:
\begin{itemize}
\item \textbf{RQ1:} What features influence the interconnections between organizations within the acquisition network?
\item \textbf{RQ2:} How has the structure of the acquisition network evolved over time?
\item \textbf{RQ3:} Which cities or countries serve as global economic hubs in terms of acquisitions?
\end{itemize}

Our contributions can be summarized as follows:
\begin{enumerate}
\item We construct the company acquisition bipartite network and its projections on acquirers and acquirees, providing their fundamental topological features.
\item We calculate and interpret structural features such as transitivity, average clustering coefficient, and the size of the largest connected components for these networks.
\item We identify the most central companies within the acquisition network and its projections by calculating various centrality measures to pinpoint influential companies in acquisition transactions.
\item We compute assortativity coefficients of different nodal attributes within the networks to assess the tendency of companies in making connections with similar companies.
\item We build exponential random graph models (ERGM) for our networks to identify which features cause connections between organizations within the networks.
\item We calculate different structural features of the acquisition network over time to provide a perspective on the evolution of the network.
\item We build the cross-city acquisition network and the cross-border acquisition network, calculating various centrality metrics to identify economic hubs among cities and countries.
\end{enumerate}

The rest of the paper is structured as follows. In the "Related work" section, we explore previous studies based on Crunchbase data and juxtapose their scope and perspectives with ours. In the "Methods and materials" section, we provide an explanation of our data and its features, along with the definition of the constructed network and metrics that we calculate in this study. In the "Results" section, we describe our analyses and findings. Following this, we answer our research questions and discuss the implications, limitations, and future directions of our study in the "Discussion" section. Finally, we conclude the paper in the "Conclusion" section.

\section{Related work}\label{sec2}

Different economic and managerial studies have utilized Crunchbase data since its foundation in 2007 to gain insights into company interactions, market trends, and strategic behaviors \cite{dalle2017using, den2020crunchbase}. They have employed various perspectives, focused on different sectors, geographic areas and time brackets, and utilized a wide range of data analysis techniques to address their research inquiries. Several studies have utilized Crunchbase data to forecast firms' success using various methodologies, including machine learning and network analysis, with different sets of features such as demographic, geographical, and fundamental company information obtained from Crunchbase or LinkedIn, as well as industry-specific features as predictors \cite{zbikowski2021machine, te2023making, kim2023succeed, bonaventura2020predicting}. Some studies employed network analysis techniques to examine investor funding or investment behavior \cite{liang2016predicting, santana2017investor, zeng2016investment, holicka2022global}. Others have concentrated on analyzing market trends within specific industries like software, healthcare, manufacturing, and FinTech \cite{cukier2016software, halminen2019factors, ferras2019new, eickhoff2017fintechs, cong2021cloud}.

However, only a few studies have sought to examine mergers and acquisition patterns using Crunchbase data. In \cite{duenas2017spatio}, the authors constructed temporal networks representing international mergers and acquisitions from this dataset, analyzing both the topological and geographical characteristics of these networks. Additionally, they conducted a comparative analysis between these networks and the international trade network, aiming to discern differences and similarities, thus gaining valuable insights. Guo et al. \cite{guo2019statistical} concentrated on Chinese companies while investigating merger and acquisition networks. They computed various centrality measures to scrutinize M\&A behaviors and constructed multiple linear regression models to evaluate the influence of diverse structural features on merger and acquisition activities. In \cite{lee2023identifying}, Lee and Geum classified merger and acquisition patterns among startups using data analysis techniques like K-means clustering and visualization. They identified five distinct clusters for M\&A activities within the Crunchbase dataset. The authors of \cite{arsini2023prediction} integrated Crunchbase data on mergers and acquisitions with other sources such as Zephyr, predicting future acquisitions through the application of machine learning and network algorithms.

To the best of our knowledge, our work is the first study on the network perspective of acquisitions in which acquiring firms and startups are analyzed geographically, determining economic hubs around the world. Additionally, unlike prior studies, we do not confine our focus to a specific time bracket, leading to a more holistic view of the acquisition patterns. 

\section{Methods and materials}\label{sec3}

\subsection{Data}
\begin{figure}[ht]
\centering
  \includegraphics[width=10cm,
  keepaspectratio]{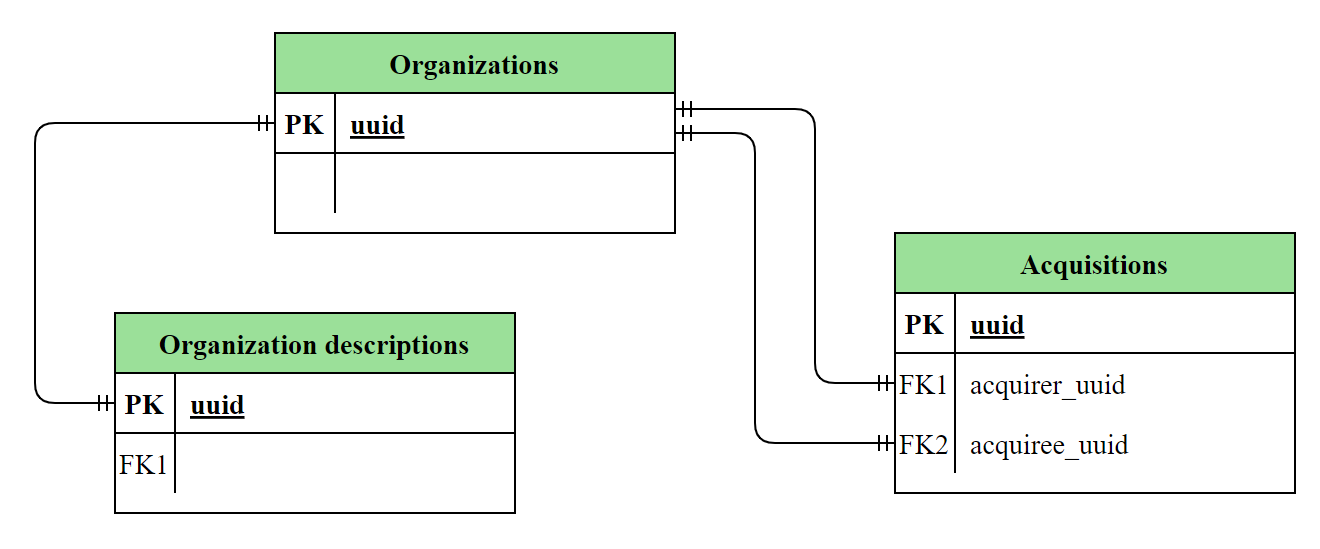}
\caption{Entity-relationship diagram of the data, comprising three tables: organizations, acquisitions, and organization descriptions. This figure shows the relationships between these entities and how they interact within the database.}
\label{fig:ERD}
\end{figure}

Figure \ref{fig:ERD} illustrates a simplified entity relationship diagram showcasing the Crunchbase data collected from the Crunchbase website by the end of July 2023. Crunchbase is a company that provides information about acquisitions, investments, and funding activities of startups. Within this comprehensive dataset, our focus was on three specific tables: organizations, acquisitions, and organization descriptions. These tables were merged based on their unique identifiers.

The organizations table contains various details about companies, including geographic features such as country, region, city, and industry features like category groups list, category list, and social network URLs such as Facebook, Twitter, LinkedIn, along with founding dates. The acquisitions table holds information related to acquisition transactions, encompassing details like acquirer ID, acquiree ID, acquisition type, and acquisition date. Lastly, the organization descriptions table provides information taken from the companies' homepages on Crunchbase.

\FloatBarrier

\subsubsection{Data preprocessing}
To prepare the data for our analyses, we take the follwing steps. First, we remove acquisitions where the date is unspecified or the type is not labeled as an acquisition. Next, we eliminate duplicate acquisitions, ensuring that only one acquisition remains for each pair of acquirer and acquiree on a specific date. Subsequently, we exclude organizations from the acquisitions if details such as their country, region, city, description, category groups list, category list, or founding date are unspecified. To ensure features are in a single-valued format, columns with list values are converted into singular values. To achieve this, we preprocess the descriptions and categories provided for each organization. We then calculate the cosine similarity between the description and each category to identify the primary category of that organization. The same process is applied to category groups. It is important to note that the categories and category groups in the initial data were sorted alphabetically, making it impossible to infer the primary ones without utilizing NLP techniques. Finally, date columns are transformed into the year-month format and then converted into numerical format for comparison across different dates. After these steps, we acquired complete data on 92,788 acquisitions, involving 119,263 organizations, out of the initial 152,923 acquisitions present in the raw data. In essence, 39.32\% of the initial data was removed during the data cleansing stage.

\subsubsection{Data exploration}

\begin{figure}[ht]
\centering
  \includegraphics[width=15cm,
  keepaspectratio]{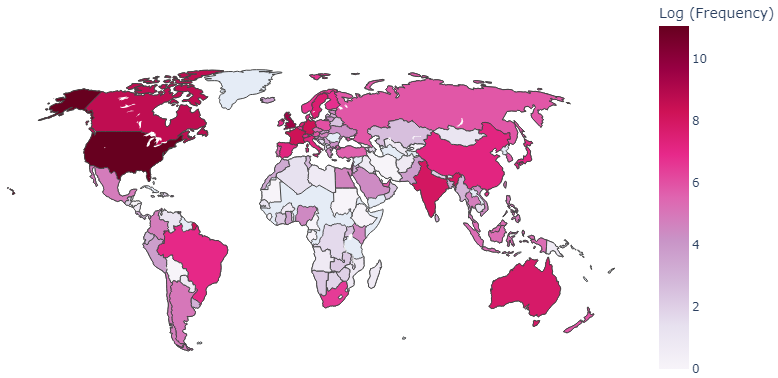}
\caption{Representation of company distribution across countries on a world map. Since the frequencies of companies follow a power law distribution, we used a logarithmic transformation to ensure that significant countries are highlighted by their colors. This figure illustrates that Western countries have a much greater number of startups compared to the rest of the world.}
\label{fig:countryMapPlot}
\end{figure}

Figure \ref{fig:countryMapPlot} depicts the global distribution of companies across different countries. The United States, United Kingdom, Canada, Germany, and France stand out with the highest number of startups in comparison to other countries. This finding corroborates earlier studies suggesting that Western nations typically exhibit higher rates of entrepreneurial activity than others \cite{fairlie2012kauffman}. Bürgel et al. \cite{burgel2001rapid} also highlight the substantial internationalization of sales among British and German startups. Furthermore, Klapper \cite{klapper2013entrepreneurship} notes a significant growth in the number of startups in France in 2009.

\FloatBarrier

\begin{figure}[ht]
\centering
  \includegraphics[width=15cm,
  keepaspectratio]{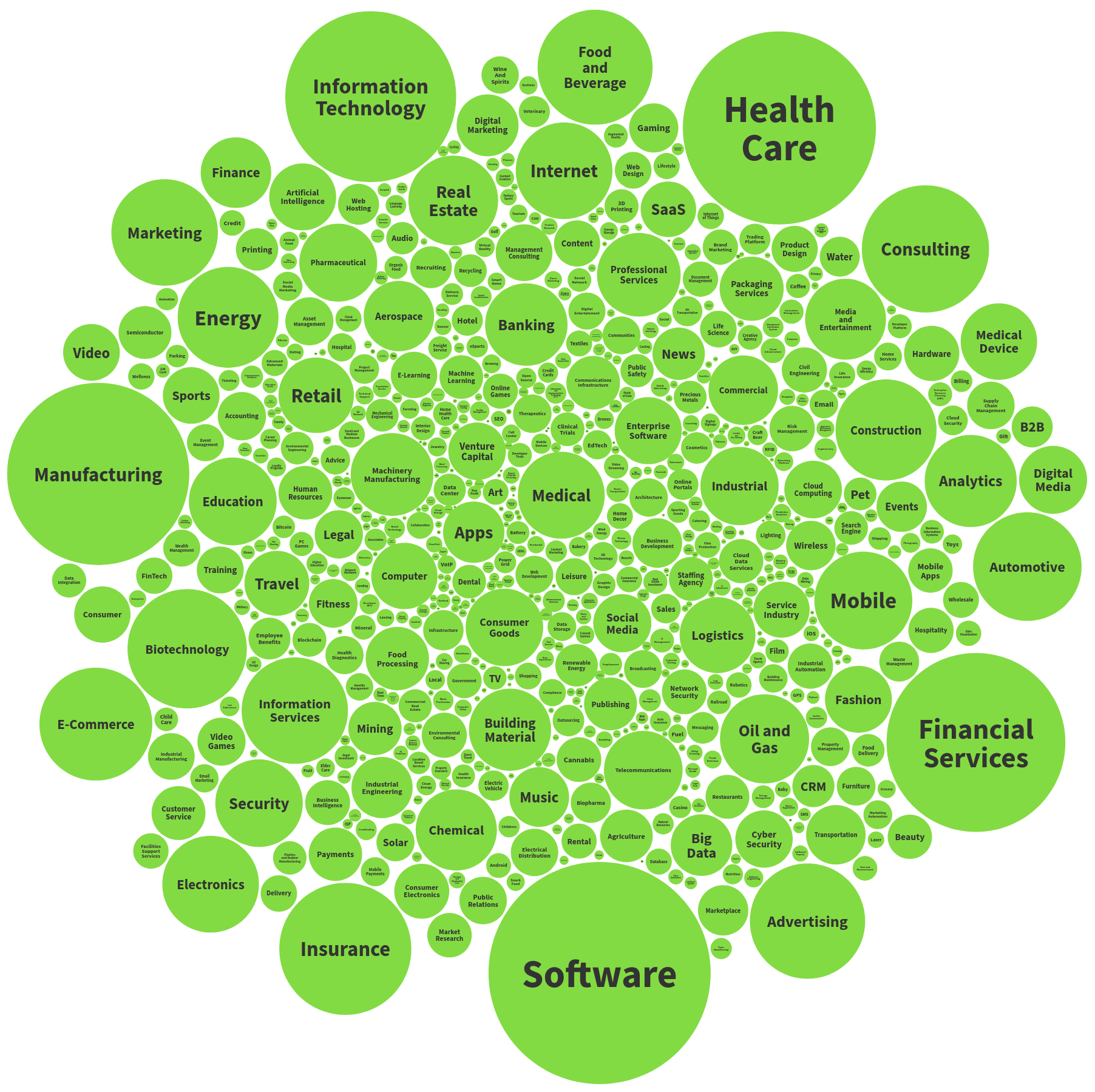}
\caption{Bubble plot of companies’ primary categories. The size of each bubble and its label corresponds to the frequency of its category. This figure demonstrates that Software, Healthcare, and Manufacturing are the prominent categories among startups.}
\label{fig:catPlot}
\end{figure}

Figure \ref{fig:catPlot} offers an overview of the primary categories within companies. Our dataset comprises 698 unique primary categories, prompting us to employ a bubble chart for visualization. This chart emphasizes categories with the highest frequencies, with each bubble's size directly proportional to its frequency within companies. Notably, Software, Health Care, Manufacturing, Financial Services, and Information Technology emerge as the categories with the highest frequencies. This finding is in the line of prior work \cite{lee2023identifying}.

\FloatBarrier

\begin{figure}[ht]
\centering
  \includegraphics[width=16cm,
  keepaspectratio]{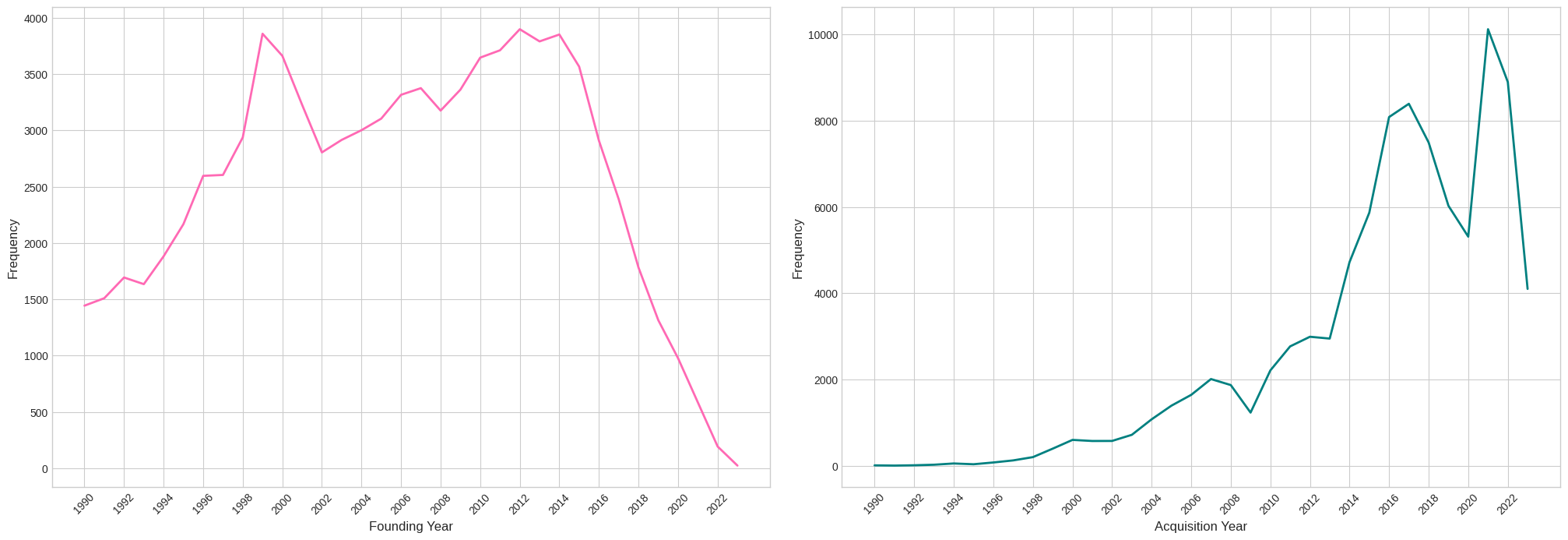}
\caption{Distributions of companies' founding years and acquisitions across different years starting from 1990. This figure illustrates that these distributions are affected by worldwide events such as the dot-com boom and the COVID-19 pandemic.}
\label{fig:yearPlots}
\end{figure}

Figure \ref{fig:yearPlots} illustrates the frequency of companies' founding years and acquisition years, spanning from 1990. The distribution of companies based on their founding years demonstrates an increasing pattern, leading to a significant initial peak in 1999. This surge may be associated with the dot-com boom in the late 1990s, contributing to a rise in technology-driven startups \cite{bhatt2022entrepreneurship}. Following this, there is a distinct decrease in 2002, succeeded by a resurgence leading to a second peak in 2012. However, post-2014, there is a discernible decline. The decrease in 2002 could be related to the aftermath of the dot-com boom, which resulted in financial problems for internet-based companies \cite{hwang2006lessons}. 

According to Figure \ref{fig:yearPlots}, the distribution of acquisitions across different years exhibits a left-skewed pattern, with a gradual increase in acquisitions followed by a steep upward trend starting from 2000. This increasing pattern could also be due to the dot-com crash. Ultimately, the highest peak in 2021 can be linked to businesses recovering post the COVID-19 pandemic, resulting in a rise in both acquisitions and investments \cite{kooli2021impact}.

\FloatBarrier

\subsection{Building networks}
\begin{figure}[ht]
\centering
  \includegraphics[width=10cm,
  keepaspectratio]{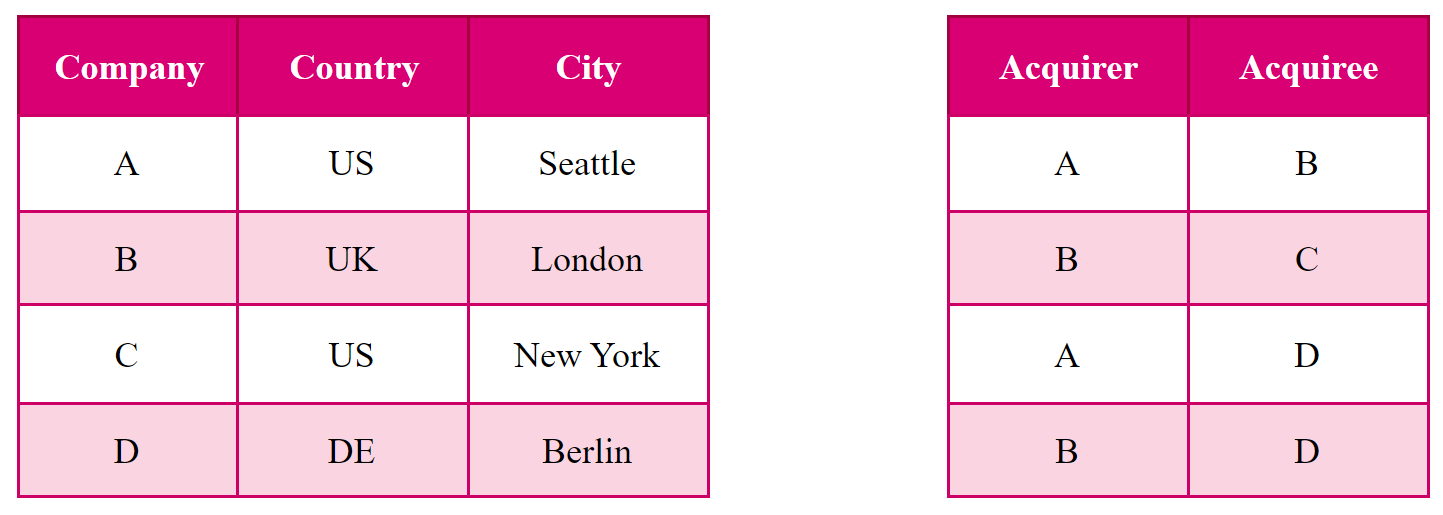}
\caption{An example of data comprising companies' countries, cities, and their acquisition relationships.}
\label{fig:dataTables}
\end{figure}

To illustrate how we construct each network, we consider a sample of the data provided in Figure \ref{fig:dataTables}.

\FloatBarrier

\subsubsection{Acquisition network}
In this network, nodes represent companies engaged in acquisitions. We establish directed edges from company $u$ to company $v$ if $u$ has acquired $v$, resulting in a directed network.

\subsubsection{Common acquirer network}
To construct this network, we establish an edge between two companies whenever a common entity has acquired both of them. This results in an undirected weighted network, where the weight of an edge signifies the number of shared acquirers between two companies.

\subsection{Common acquiree network}
To build this network, we establish an undirected edge between two companies if they have acquired the same company. The weight of these edges corresponds to the number of shared acquirees between the two companies. This process results in an undirected weighted network.

\begin{figure}[ht]
\centering
  \includegraphics[width=12cm,
  keepaspectratio]{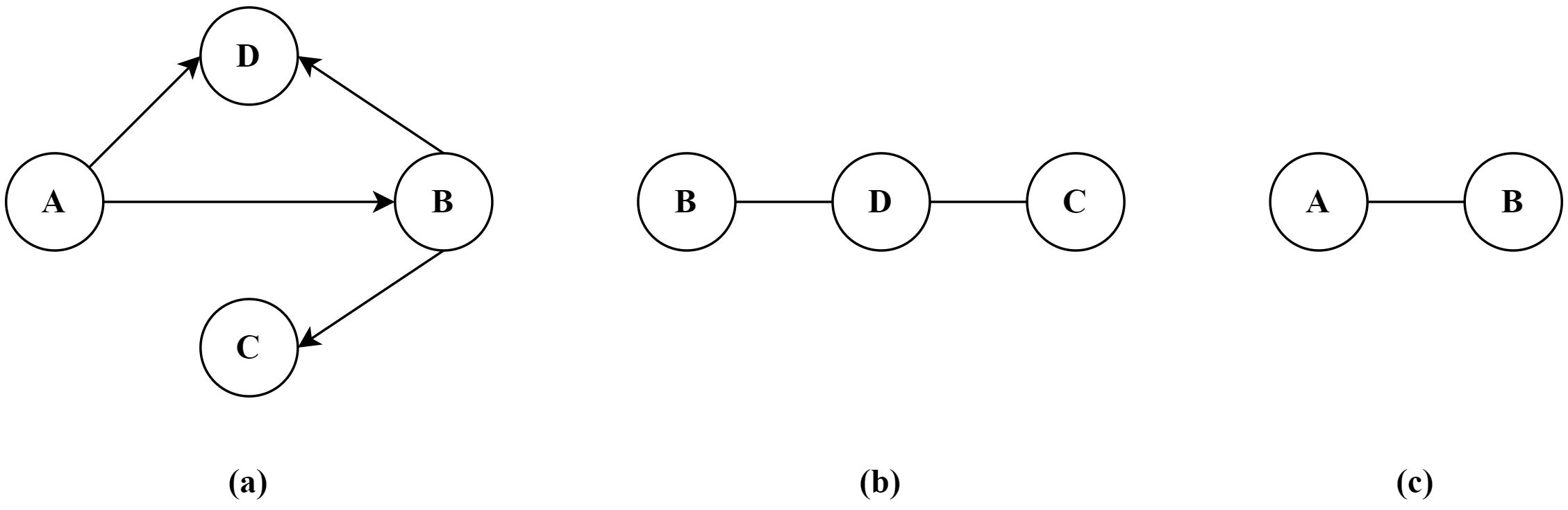}
\caption{Networks based on the data in Figure \ref{fig:dataTables}: \textbf{a} Acquisition network, \textbf{b} Common acquirer network, \textbf{c} Common acquiree network}
\label{fig:companyNets}
\end{figure}

Figure \ref{fig:companyNets} represents examples of the acquisition network, common acquirer network, and common acquiree network according to the data in Figure \ref{fig:dataTables}.

\FloatBarrier

\subsubsection{Cross-city acquisition network}
In the cross-city acquisition network, nodes represent cities of companies, and directed edges connect city $u$ to city $v$ if a company from city $u$ has acquired a company in city $v$. The weight of each edge represents the number of acquisitions between the two cities.

\subsubsection{Cross-border acquisition network}
To construct the cross-border acquisition network, we establish a directed edge from country $u$ to country $v$ whenever a company in $u$ acquires a company in $v$. This creates a weighted directed network where edge weights signify the number of acquisitions between countries.

\begin{figure}[ht]
\centering
  \includegraphics[width=10cm,
  keepaspectratio]{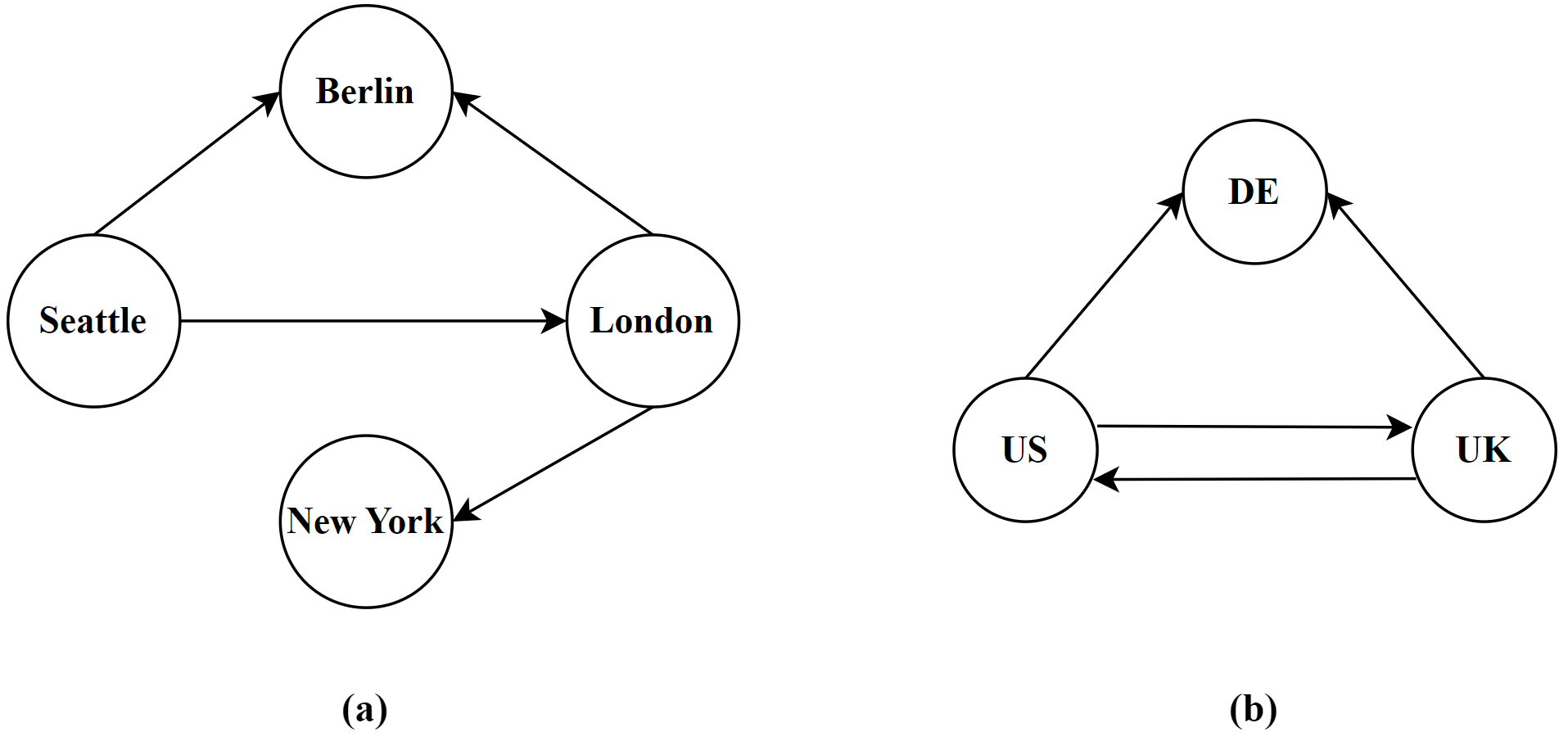}
\caption{Networks based on the data in Figure \ref{fig:dataTables}: \textbf{a} Cross-city acquisition network, \textbf{b} Cross-border acquisition network}
\label{fig:geoNets}
\end{figure}

Figure \ref{fig:geoNets} shows examples of the cross-city acquisition network and cross-border acquisition network, derived from the data in Figure \ref{fig:dataTables}.

\FloatBarrier

\subsection{Metrics}

\subsubsection{Transitivity}
We measure the transitivity to assess the strength of the community structures of the acquisition network and its projections. This metric aligns with our research as it identifies the level of cohesion and resilience in the network, which are useful for making acquisition decisions. The transitivity of a network is calculated as the proportion of all potential triangles existing within the network. The formula for transitivity is:
\begin{eqnarray}
\frac{(number\: of\: triangles \times 3)}{(number\: of\: connected\: triples)}
\end{eqnarray}

as proved by \cite{Newman2010}. 

\subsubsection{Average clustering coefficient}
We measure the average clustering coefficient to assess the level of interconnectedness and community structure within the acquisition network and its projections. This metric is relevant to our study as it specifies the degree to which companies tend to create tightly-knit groups within the network, providing a better understanding of market dynamics. The formula for the average clustering coefficient is as follows:
\begin{eqnarray}
   C = \frac{1}{n}\sum_{v \in G}c_v,
\end{eqnarray}

where $n$ is the number of nodes in the network, and $c_v$ is the local clustering coefficient of node $v$ \cite{kaiser2008mean}. 

\subsubsection{Average shortest path length}
We measure the average shortest path length for the largest connected component of the acquisition network and its projections to evaluate the mean distance between each pair of companies. This metric is useful for our research as it evaluates the connectivity of the network and the speed at which market trends and economic influences can propagate through it. The average shortest path length of a network is computed as the average length of the shortest paths between all pairs of nodes in the network. The formula for average shortest path length is as follows:
\begin{eqnarray}
   S = \frac{\sum_{u, v}d(u, v)}{n \times (n-1)},
\end{eqnarray}

where $d(v, u)$ is the length of shortest path between nodes $v$ and $u$, and $n$ is the number of nodes in the network \cite{Newman2010}.

\subsubsection{Assortativity}
We measure the assortativity for various node attributes to identify which attributes are more influential in shaping the connections within the acquisition network and its projections. This metric captures the tendency among companies to engage in acquisition transactions with other companies that have a similar set of attributes, making it fitting for our research. The assortativity coefficient of a network calculates the proportion of connections in the network that occur between nodes of the same type based on a specific attribute \cite{Newman2010}. 

\subsubsection{Betweenness centrality}
We measure the betweenness centrality for the nodes in the cross-city acquisition network to determine which cities are influential in terms of their control over acquisition transactions between other cities. Betweenness centrality is helpful for this purpose as it identifies cities that act as intermediaries between different cities within the network, which can be targeted for strategic acquisitions. This metric calculates the extent to which a node lies on paths between other nodes in a network. The formula for betweenness centrality is as follows:
\begin{eqnarray}
   B_u = \sum_{i, j}\frac{\sigma(i, u, j)}{\sigma(i, j)},
\end{eqnarray}

where $\sigma(i, u, j)$ represents the number of shortest paths between nodes $i$ and $j$ that pass through node or edge $u$, and $\sigma(i, j)$ is the total number of shortest paths between $i$ and $j$. The sum is computed over all pairs $i$, $j$ of distinct nodes in the network \cite{Newman2010}. 

\subsubsection{Closeness centrality}
We measure the closeness centrality to determine which cities are closer to other cities in the cross-city acquisition network. Closeness centrality is useful for our analysis as cities with high values for this measure can swiftly affect economic trends and business practices, and facilitate acquisition. This metric calculates the mean distance from a node to other nodes in a network. The formula for closeness centrality is as follows:
\begin{eqnarray}
   C_u = \frac{n-1}{\sum_{v=1}^{n-1}d(v, u)},
\end{eqnarray}

where n indicates the count of nodes accessible from node $u$, and $d(v, u)$ represents the shortest distance between nodes $v$ and $u$ \cite{Newman2010}.

\subsubsection{Weighted degree centrality}
We measure the weighted degree centrality to assess which cities acquire a greater number of organizations from other cities in the cross-city acquisition network. Weighted degree centrality is particularly suitable for this research as it offers a more nuanced measure of a city's influence by not only considering the number of its connections but also their strength. This metric calculates the sum of the weights of the edges connected to a node in a network. The formula for weighted degree centrality is as follows:
\begin{eqnarray}
   WD_u = \sum_{v}w(v, u),
\end{eqnarray}

where node $v$ is a neighbor of node u, and $w(v, u)$ is the weight of the edge between nodes $v$ and $u$ \cite{Newman2010}. 

\subsubsection{PageRank centrality}
We measure the PageRank centrality to identify influential cities in the cross-city acquisition network in terms of their incoming link counts. PageRank centrality is especially fitting for our analysis since it not only considers the number of incoming links but also reflects the significance of the linking cities. This metric calculates the ranking of a node within a network by considering the incoming link structure. The formula for PageRank centrality is as follows:
\begin{eqnarray}
   x_i = \alpha \sum_{j} {A_{ij}\frac{x_j}{d_j^{out}}} + \beta,
\end{eqnarray}

where $A_{ij}$ is an element of the adjacency matrix, $d_j^{out}$ denotes the out-degree of node $j$, and $\alpha$ and $\beta$ are positive constants \cite{Newman2010}.

\subsubsection{Eigenvector centrality}
We measure the eigenvector centrality to assess the transitive influence among cities in the cross-city acquisition network. Eigenvector centrality is suitable for this study as it considers not only the number of links but also connections to other highly influential cities. This metric calculates the centrality of a node considering the centrality of its neighboring nodes. It is determined by solving the equation
\begin{eqnarray}
   Ax = \lambda x,
\end{eqnarray}

where $A$ represents the adjacency matrix of the network and $x$ is the eigenvector. The eigenvalue associated with $A$ influences this calculation \cite{Newman2010}. 

\subsubsection{Authority centrality}
We measure the authority centrality to assess the most influential countries in the cross-border acquisition network. This metric is specifically relevant to our analysis as it highlights countries that are not only well-connected but also recognized as authorities in acquisitions. The authority centrality of a node within a network is derived from the principal eigenvector of the matrix $A^TA$, where $A$ stands for the network's adjacency matrix \cite{kleinberg1999authoritative}. 

\section{Results}\label{sec4}

In this section, we provide a detailed description of the analyses performed and our findings, along with their interpretations.

\subsection{Acquisition network}
\begin{figure}[ht]
\centering
  \includegraphics[width=15cm,
  keepaspectratio]{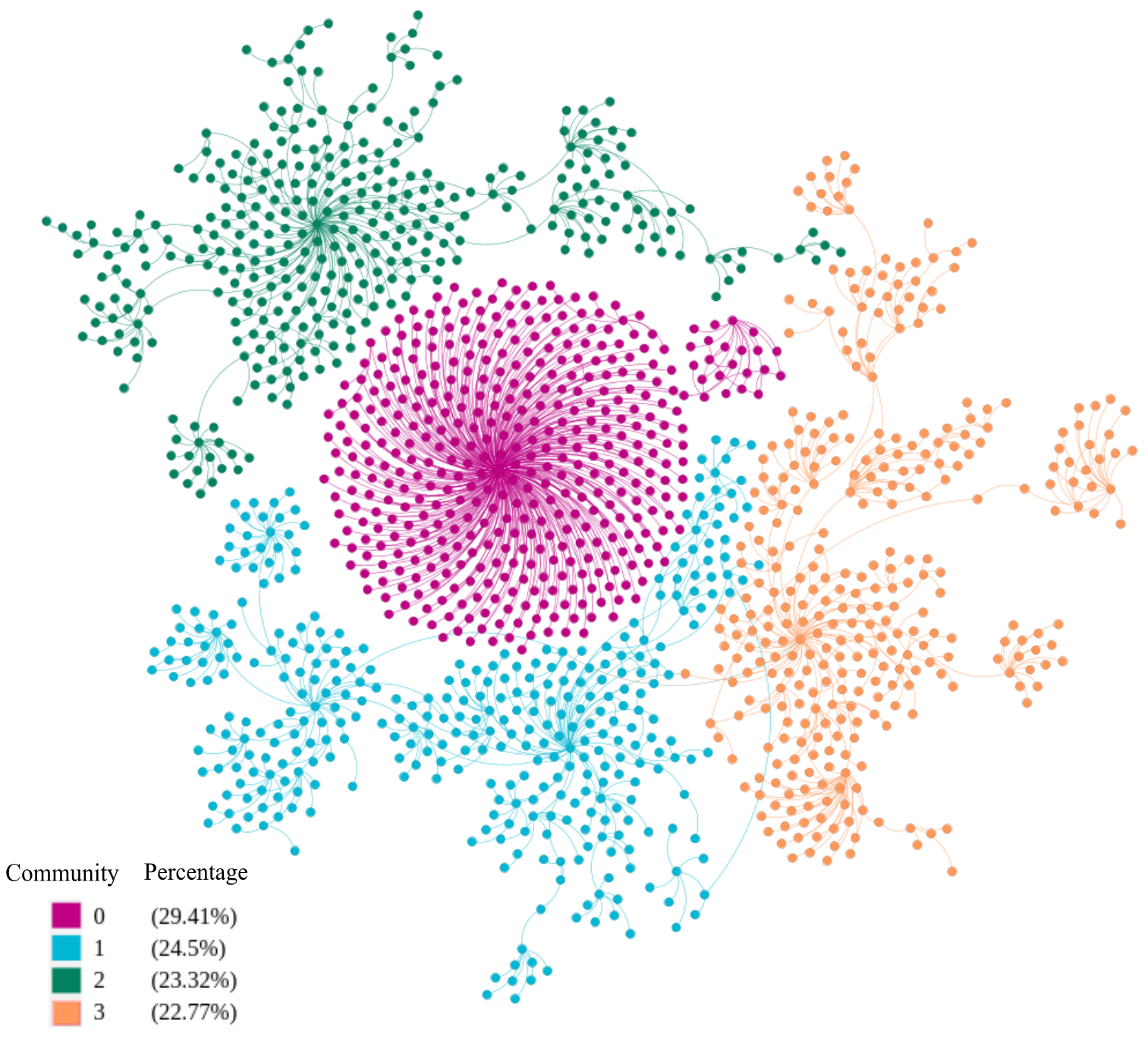}
\caption{A sampled subgraph of the acquisition network. Nodes are colored based on their community detected via the Louvain algorithm \cite{blondel2008fast}. The percentage of startups in each community is specified next to the community ID.}
\label{fig:acquisitionNet}
\end{figure}

The acquisition network comprises 119,263 nodes and 92,788 edges. Considering the network's extensive size, we selected a sample for visualization purposes. To accomplish this, we displayed the four largest Louvain communities from the largest weakly connected component of the network. Figure \ref{fig:acquisitionNet} illustrates this subgraph, encompassing 1,445 nodes and 1,447 edges, highlighting four distinct communities within the network.

\FloatBarrier

\subsection{Common acquirer network}
\begin{figure}[ht]
\centering
  \includegraphics[width=15cm,
  keepaspectratio]{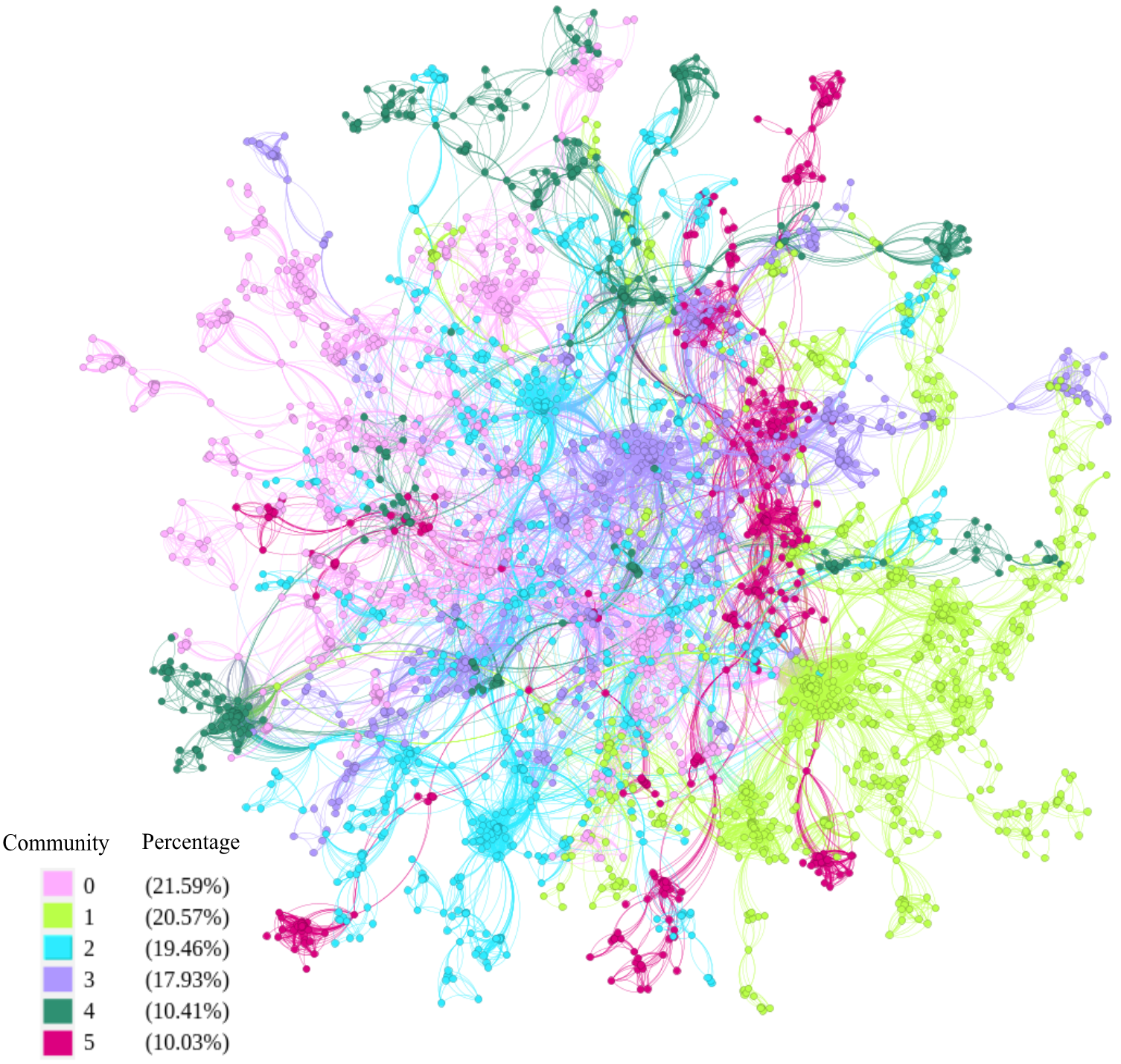}
\caption{A sampled subgraph of the common acquirer network. Nodes are colored based on their community detected via the Louvain algorithm. The percentage of startups in each community is specified next to the community ID.}
\label{fig:commonAcquirerNet}
\end{figure}

The common acquirer network encompasses 68,297 nodes and 639,064 edges. Figure \ref{fig:commonAcquirerNet} depicts a sample of this network, showcasing the six largest Louvain communities within the largest connected component of the network. The resulting sample consists of 3,937 nodes and 44,340 edges. As illustrated in Figure \ref{fig:commonAcquirerNet}, each community within the network is highlighted with a specific color. Companies within each community primarily originate from the United States. Table \ref{tab:tab1} presents the percentage of the major industry category within each community. As shown, most communities have Health Care or Manufacturing as their primary industry category, aligning with our previous observation indicating that these industries are the most common category groups among startups.

\FloatBarrier

\begin{table*}[htbp]
\small
\caption{Percentage of major industry categories within communities of the sample common acquirer network in Figure \ref{fig:commonAcquirerNet}. The table highlights that Manufacturing and Health Care are the dominant industry categories in these communities.}
\centering
\begin{tabular}{ccccc}
\hline
Community & Industry category & Portion \\
\hline
0 & Health Care & 18.35\% \\
1 & Manufacturing & 14.81\% \\
2 & Manufacturing & 15.01\% \\
3 & Manufacturing & 11.76\% \\
4 & Manufacturing & 20.48\% \\
5 & Food and Beverage & 12.40\% \\
\hline
\end{tabular}
\label{tab:tab1}
\end{table*}

\subsection{Common acquiree network}
\begin{figure}[ht]
\centering
  \includegraphics[width=15cm,
  keepaspectratio]{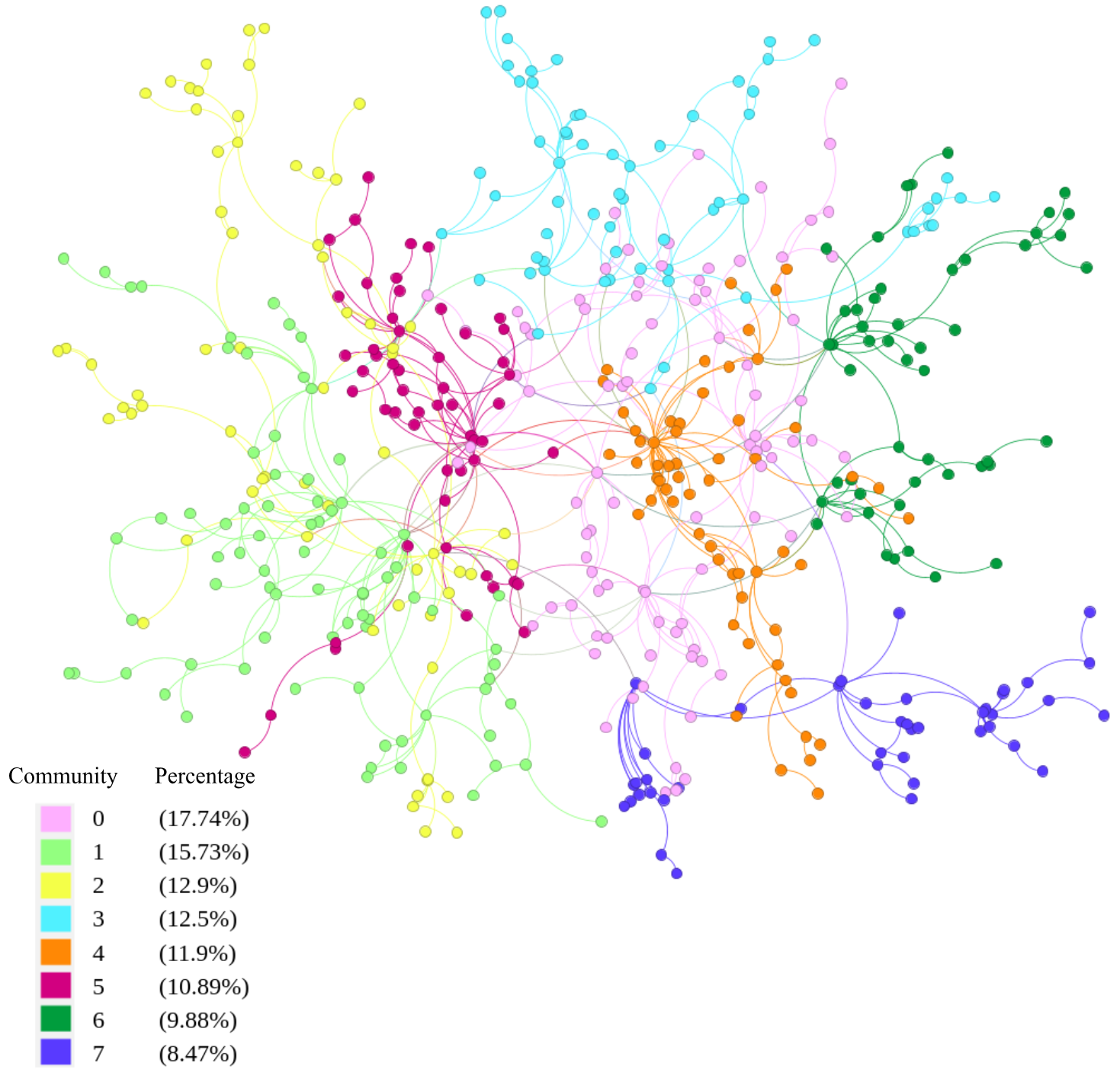}
\caption{A sampled subgraph of the common acquiree network. Nodes are colored based on their community detected via the Louvain algorithm. The percentage of companies in each community is specified next to the community ID.}
\label{fig:commonAcquireeNet}
\end{figure}

The common acquiree network comprises 4,063 nodes and 3,115 edges. To facilitate visualization, we have selected the eight largest communities from within the largest connected component, creating a sample network composed of 496 nodes and 593 edges. Figure \ref{fig:commonAcquireeNet} depicts this visualization with communities detected using the Louvain algorithm. Each community mainly comprises acquirer companies from the United States and the Financial Services industry. Table \ref{tab:tab2} represents the most common regions within each community, along with their respective percentages. According to this table, New York and England are the major regions from which acquirer companies originate, aligning with prior studies that highlight the prominence of these regions in startup ecosystems \cite{mulas2016new, hart2009economic}.

\FloatBarrier

\begin{table*}[htbp]
\small
\caption{Percentage of major regions within communities of the sample common acquiree network in Figure \ref{fig:commonAcquireeNet}. The table highlights that New York and England are the prominent regions in these communities.}
\centering
\begin{tabular}{ccccc}
\hline
Community & Region & Portion \\
\hline
0 & New York & 10.23\% \\
1 & California & 15.38\% \\
2 & England & 12.50\% \\
3 & Ontario & 11.29\% \\
4 & New York & 8.47\% \\
5 & New York & 14.81\% \\
6 & New York & 12.24\% \\
7 & England & 14.29\% \\
\hline
\end{tabular}
\label{tab:tab2}
\end{table*}

\subsection{Structural analysis}
Table \ref{tab:acquisitionStructure}, \ref{tab:commonAcquirerStructure}, and \ref{tab:commonAcquireeStructure} represent a summary of the structural properties of the acquisition network, the common acquirer network, and the common acquiree network, respectively. We considered the Erdős–Rényi (ER) model \cite{bollobas1998random} to create baseline models of each of these networks, which are useful for validating the observed topological features of the networks. These baseline models have the same number of nodes and probability of edge creation as their respective real networks. Based on the data provided in the tables, it is evident that these networks have low densities, suggesting the likelihood of organizations forming smaller, distinct communities within these networks. The transitivity of the acquisition network is remarkably low, while that of the common acquirer network is close to 1. The common acquiree network exhibits a transitivity that is relatively low, but it is higher compared to that of the acquisition network. This outcome was anticipated since projection networks typically manifest strong community structures, resulting in high transitivity \cite{orman2013empirical}. 

We computed the average clustering coefficient and average shortest path length within the largest weakly connected component of the acquisition network. These metrics indicate that this network is not well-connected. The low average clustering coefficient suggests the presence of weak ties, and on average, it takes at least 14 steps to reach from one randomly selected organization to another. Conversely, the common acquirer network exhibits a contrasting scenario. It boasts a high average clustering coefficient and a small average shortest path length. This pattern aligns with the network's configuration of well-connected clusters of organizations sharing common acquirers, as anticipated. On the other hand, the common acquiree network has a moderate average clustering coefficient and an average shortest path length close to that of the common acquirer network. The transitivity and average clustering coefficient values of the random models are zero or close to zero, which is expected, as ER models cannot capture these common features of real networks \cite{burda2004network}. The average shortest path lengths of the ER models for the acquisition network and the common acquiree network are higher than those of the networks themselves, indicating the presence of economic hubs that shorten the distance between companies. However, for the common acquirer network, the ER model exhibits a lower average shortest path length, which is in line with the fragmentation of the network into distinct clusters with fewer connections between them.

Additionally, we calculated the sizes of the largest connected components (both weakly and strongly connected for the acquisition network) in these networks. Notably, the presence of a strongly connected component with a size of three in the acquisition network indicates that at most three organizations are mutually involved in acquisitions. Conversely, the substantial size of the largest connected component (and the largest weakly connected component for the acquisition network) underscores the robust and centralized structures inherent in these networks. The size of the largest strongly connected component of the ER model for the acquisition network is 5, which is higher than that of the real network. Moreover, in terms of the largest weakly connected component size, all the three networks exhibits lower values compared to their ER counterparts. These observations are expected as random models do not have community structure and their degree distributions are more uniform, resulting in the formation of large connected components. 

\begin{table*}[htbp]
\small
\caption{Comparison of structural properties between the acquisition network and its baseline model. This table reveals that the acquisition network has a decentralized structure comprising smaller, distinct communities.}
\centering
\begin{tabular}{ccccc}
\hline
\multirow{3}{*}{Property} & \multicolumn{2}{c}{Measurement} \\
& Acquisition network & Baseline model \\
\hline
Density & $6.524 \times 10^{-6}$ & $6.530 \times 10^{-6}$ \\
Transitivity & $6.024 \times 10^{-5}$ & $2.738 \times 10^{-5}$ \\
\makecell{Average clustering \\ coefficient} & 0.002 & 0 \\
\makecell{Average shortest path \\ length} & 14.194 & 23.1727 \\
\makecell{Largest weakly connected \\ component size} & 25285 & 73669 \\
\makecell{Largest strongly connected \\ component size} & 3 & 5 \\
\hline
\\
\end{tabular}
\label{tab:acquisitionStructure}

\caption{Comparison of structural properties of the common acquirer network and its baseline model. This table illustrates that the common acquirer network has tighter interconnections.}
\centering
\begin{tabular}{ccccc}
\hline
\multirow{3}{*}{Property} & \multicolumn{2}{c}{Measurement} \\
& Common acquirer network & Baseline model\\
\hline
Density & $2.740 \times 10^{-4}$ & $2.740 \times 10^{-4}$ \\
Transitivity & 0.993 & $2.707 \times 10^{-4}$ \\
\makecell{Average clustering \\ coefficient} & 0.959 & $2.695 \times 10^{-4}$ \\
\makecell{Average shortest path \\ length} & 7.217 & 4.0775 \\
\makecell{Largest connected \\ component size} & 10427 & 68297 \\
\hline
\\
\end{tabular}
\label{tab:commonAcquirerStructure}

\caption{Comparison of structural properties of the common acquiree network and its baseline model. This table indicates that the common acquiree network exhibits more closely connected nodes.}
\centering
\begin{tabular}{ccccc}
\hline
\multirow{3}{*}{Property} & \multicolumn{2}{c}{Measurement} \\
& Common acquiree network & Baseline model\\
\hline
Density & $3.775 \times 10^{-4}$ & $3.890 \times 10^{-4}$ \\
Transitivity & 0.081 & $5.896  \times 10^{-4}$ \\
\makecell{Average clustering \\ coefficient} & 0.108 & $3.216 \times 10^{-4}$ \\
\makecell{Average shortest path \\ length} & 6.922 & 15.451 \\
\makecell{Largest connected \\ component size} & 1098 & 2591 \\
\hline
\end{tabular}
\label{tab:commonAcquireeStructure}
\end{table*}
\FloatBarrier

We also identify the most central nodes within these networks to pinpoint hub companies in acquisitions. Table \ref{tab:tab4} presents the most central companies in the networks based on different centrality measures, providing a holistic view of these economic hubs. According to our results, Gallagher, HUB International, Napster, Solidscape, and Hume Brophy emerge as the most centric acquiring firms in the acquisition network, most of which are from the United States and have already been acquired at the present time. In the common acquirer network, William H. Connolly \& Co., Raet, Nets, and Webhelp stand out as the most centric acquired firms. Most of them are located in European countries, including the Netherlands, Denmark, and France, emphasizing the significance of European startups in acquisition transactions \cite{pisoni2018startups}. In the common acquiree network, Advent International has the highest values for all centrality metrics, highlighting its dominant role as a private equity firm based in the United States.

\begin{table*}[htbp]
\small
\caption{The most central nodes within the acquisition network and its projections. This table demonstrates that pivotal acquiring firms are from the United States, while European startups are central in acquisition transactions.}
\centering
\begin{tabular}{ccccc}
\hline
\multirow{3}{*}{Centrality metric} & \multicolumn{3}{c}{Most central nodes} \\
& Acquisition network & \makecell{Common acquirer \\ network} & \makecell{Common acquiree \\ network} \\
\hline
Degree centrality & Gallagher & William H. Connolly \& Co. & Advent International \\
Betweenness centrality & HUB International & Raet & Advent International \\
Closeness centrality & Napster & Nets & Advent International \\
Eigenvector centrality & Solidscape & William H. Connolly \& Co. & Advent International \\
PageRank centrality & Hume Brophy & Webhelp & Advent International \\
\hline
\end{tabular}
\label{tab:tab4}
\end{table*}

\subsection{Assortativity analysis}
Table \ref{tab:tab5} provides an overview of assortativity coefficients corresponding to various node features in the acquisition network, common acquirer network, and common acquiree network. As shown, the acquisition network exhibits the highest assortativity coefficient with respect to the country feature. This suggests that organizations are more likely to pursue acquisitions within their own country, consistent with previous studies that highlight a higher inclination among companies to engage in domestic acquisitions \cite{ciuchta2018buy}. Such acquisitions are known to yield greater returns for acquirers \cite{uysal2008geography}. The next feature with the highest assortativity is the category group, indicating organizations' preference for acquiring those in the same industries as themselves, aligning with prior studies \cite{schildt2006buys, lee2023identifying}. The third highest coefficient corresponds to the founding date of the organizations, suggesting that organizations involved in acquisitions tend to have close founding dates, consistent with previous research \cite{lee2023identifying}. Notably, all coefficients are positive, indicating a prevailing inclination among organizations to acquire entities similar to themselves, aligning with findings from prior studies \cite{schildt2006buys}.

In the common acquirer network, the highest assortativity is observed for the node degree, as anticipated. This is due to the fact that 96.87\% of acquiree organizations, based on the indegree of nodes in the acquisition network, have only one acquirer. Consequently, neighbors in the common acquirer network tend to have similar node degrees. The second and third highest assortativity values are linked to the category group and founding date, respectively. Consistent with previous studies \cite{kennedy2002matching,ransbotham2010target}, this finding suggests a propensity among acquirers to target organizations within specific industries or age brackets.

Similar to the acquisition network, the highest assortativity coefficients in the common acquiree network belong to the country, category group, and founding date, respectively. In essence, organizations from the same country, industry, or age bracket are more inclined to acquire similar organizations.

\begin{table*}[htbp]
\small
\caption{Assortativity coefficients of the acquisition network and its projections. The data in this table suggest that companies from the same country or category, or with similar founding dates, tend to engage in acquisition transactions with each other.}
\centering
\begin{tabular}{ccccc}
\hline
\multirow{3}{*}{Feature} & \multicolumn{3}{c}{Assortativity coefficient} \\
& Acquisition network & \makecell{Common acquirer \\ network} & \makecell{Common acquiree \\ network} \\
\hline
Country & 0.472 & 0.216 & 0.286 \\
Region & 0.167 & 0.042 & 0.083 \\
City & 0.088 & 0.013 & 0.054 \\
Category group & 0.292 & 0.236 & 0.157 \\
Category & 0.177 & 0.162 & 0.082 \\
\makecell{Founding date \\ (year-month)} & 0.248 & 0.228 & 0.097\\
Node degree & 0.019 & 0.986 & 0.088 \\
\hline
\end{tabular}
\label{tab:tab5}
\end{table*}

\subsection{ERGM analysis}
Our ERGM models for the acquisition network, common acquirer network, and common acquiree network utilize similar sets of node features and “edges” as parameters. Table \ref{tab:tab6} provides a summary of the ERGM statistics. According to the results, for these networks, all predictors are statistically significant $(p < 0.0001)$ with positive estimates, except for the founding date, indicating that organizations with similar values for each feature are more likely to be connected. For instance, organizations within the same country are more likely to share the acquirer-acquiree relationship. This aligns with prior research indicating that companies tend to purchase firms located in closer geographic proximity \cite{erel2011determinants}. Conversely, negative estimations regarding the founding date suggest that this feature cannot be the cause of the connection between organizations within the networks. Furthermore, the acquisition network and the common acquirer network have positive estimates for “edges”, indicating that they have higher densities than random networks with similar numbers of nodes. In other words, nodes in both networks are more inclined to be connected to each other \cite{lusher2013exponential}. The negative estimate for “edges” indicates that the common acquiree network exhibits lower density in comparison to random networks possessing similar topological features. The Akaike Information Criterion (AIC) values for each model, as shown in Table \ref{tab:tab6}, reveal that the ERGM of the common acquiree network has a better fit (lower AIC value) compared to other networks.

\begin{table*}[htbp]
\small
\caption{Estimates from the ERGMs of the acquisition network and its projections. The data in this table show that companies with the same set of features, except for the founding date, are more likely to be connected to each other in these networks.}
\centering
\begin{tabular}{ccccc}
\hline
\multirow{3}{*}{Variable} & \multicolumn{3}{c}{Estimate (standard error)} \\
& Acquisition network & \makecell{Common acquirer \\ network} & \makecell{Common acquiree \\ network} \\
\hline
edges & \makecell{8.135*** \\ $(2.005 \times 10^{-1})$} & \makecell{9.803*** \\ $(1.053 \times 10^{-1})$} & \makecell{-4.634*** \\ $(1.115)$} \\
Country & \makecell{1.161*** \\ $(7.342 \times 10^{-3})$} & \makecell{0.721*** \\ $(2.605 \times 10^{-3})$} & \makecell{0.787*** \\ $(3.840 \times 10^{-2})$} \\
Region & \makecell{1.117*** \\ $(1.169 \times 10^{-2})$} & \makecell{0.544*** \\ $(5.706 \times 10^{-3})$} & \makecell{0.503*** \\ $(8.354 \times 10^{-2})$} \\
City & \makecell{1.373*** \\ $(1.505 \times 10^{-2})$} & \makecell{0.479*** \\ $(1.107 \times 10^{-2})$} & \makecell{0.693*** \\ $(1.031 \times 10^{-1})$} \\
Category group & \makecell{1.794*** \\ $(9.213 \times 10^{-3})$} & \makecell{1.481*** \\ $(3.856 \times 10^{-3})$} & \makecell{1.193*** \\ $(5.116 \times 10^{-2})$} \\
Category & \makecell{1.351*** \\ $(1.104 \times 10^{-2})$} & \makecell{1.553*** \\ $(4.454 \times 10^{-3})$} & \makecell{0.882*** \\ $(7.056 \times 10^{-2})$} \\
\makecell{Founding date \\ (year-month)} & \makecell{$-5.298 \times 10^{-5}***$ \\ $(5.048 \times 10^{-7})$} & \makecell{$-4.678 \times 10^{-5}***$ \\ $(2.647 \times 10^{-7})$} & \makecell{$-9.727 \times 10^{-6}***$ \\ $(2.819 \times 10^{-6})$} \\
AIC & 2226707 & 11085347 & 53343 \\
\bottomrule
  \multicolumn{1}{c}{$* p < 0.05, ** p < 0.01, *** p < 0.001$}
\end{tabular}
\label{tab:tab6}
\end{table*}

\subsection{Temporal analysis}
To investigate the evolution of the acquisition network over time, we begin our analysis with the January 2000 network and subsequently consider one-month intervals between consecutive time points, resulting in 283 time points up to the present data collection date. At each time point, we compute various structural metrics for the corresponding acquisition network. 

\begin{figure}[ht]
\centering
  \includegraphics[width=16cm,
  keepaspectratio]{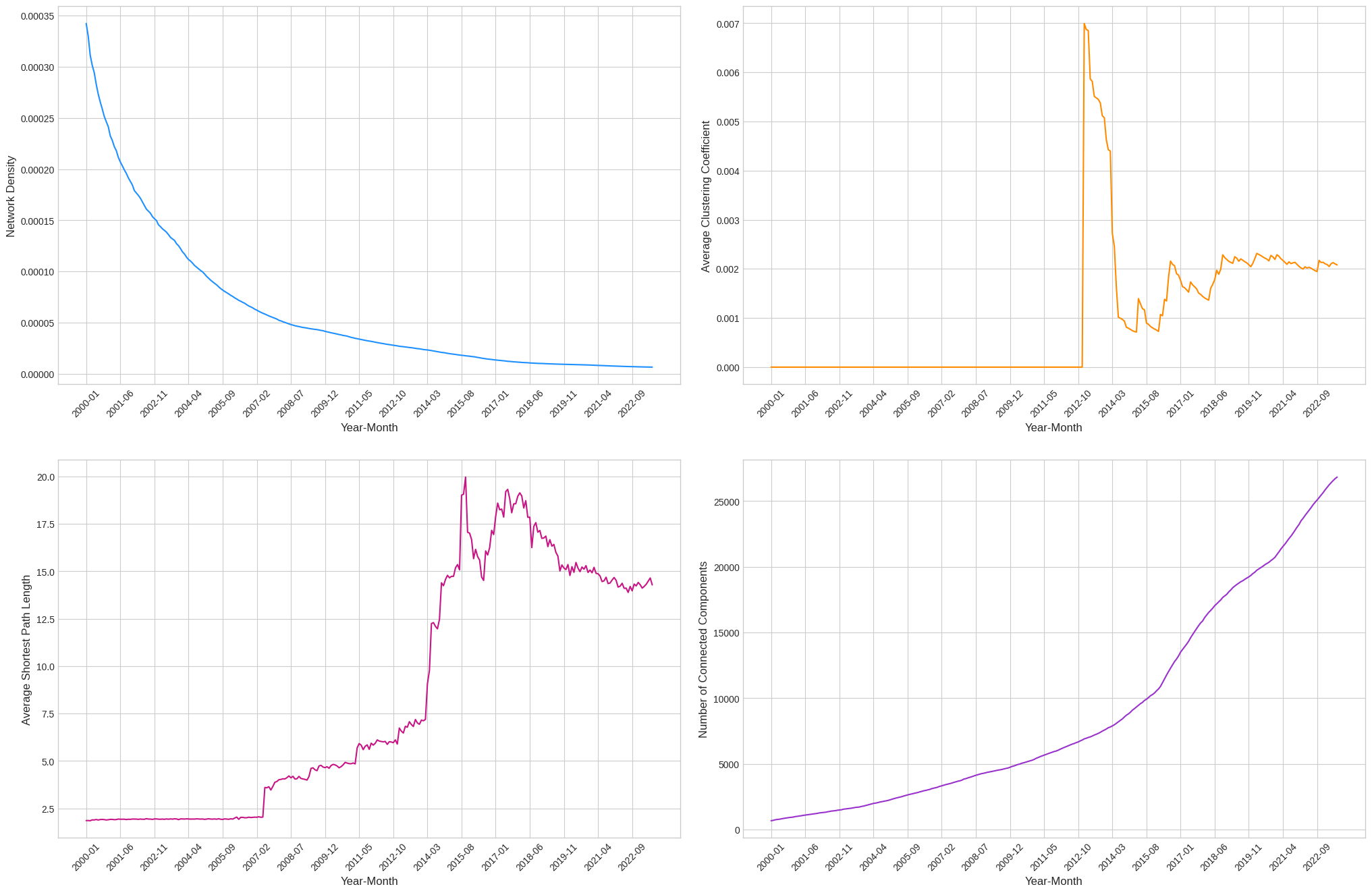}
\caption{Time series of density, average clustering coefficient, average shortest path length, and the number of weakly connected components within the acquisition network starting from 2000. These plots together suggest that the network has become more fragmented while comprising denser clusters.}
\label{fig:temporalPlots}
\end{figure}

Figure \ref{fig:temporalPlots} illustrates the temporal changes in the density, average clustering coefficient, average shortest path length, and the number of weakly connected components of the network. The visual representation reveals a declining trend for density, suggesting that the network has gradually become sparser over time. In essence, organizations within the network have exhibited reduced connections over time. This trend could be attributed to the increase in the number of investors and venture capitalists, and the fact that we are no longer observing a dominant monopoly \cite{fichtner2020rise, lerner2022venture}.

According to Figure \ref{fig:temporalPlots}, the network's average clustering coefficient remained at zero before 2012, indicating a lack of connections between the organizations' neighborhoods within the network. However, starting from 2012, it shows an upward trend, reaching its peak in 2013. Subsequently, there is a decline observed until 2014, followed by a general increasing trend with intermittent fluctuations. This pattern suggests a gradual densification of local connections over time, even as the overall connections have become sparser. It could be related to the European debt crisis in the early 2010s, which might have prompted many companies to attempt acquiring neighboring businesses as a strategic move to expand and overcome economic challenges \cite{nicholson2014impact}.

Based on this figure, the average shortest path length of the acquisition network remains around 2 until 2007, after which it steadily increases, reaching its initial peak in the same year. Subsequently, from 2007 onward, it gradually ascends to its second peak in 2015. Following this, it displays a gradual decrease and eventually fluctuates around 14. Essentially, this suggests that in recent years, on average, organizations required approximately 14 hops to reach another organization within the network. This pattern could be related to the 2007–2008 financial crisis, which might have resulted in the fragmentation of the acquisition network in the subsequent years \cite{reddy20142007}.

As illustrated in Figure \ref{fig:temporalPlots}, the number of weakly connected components follows an increasing pattern, indicating a rising level of segregation in the network. In fact, the network has become more fragmented and less cohesive, further highlighting the absence of a monopoly in acquisition transactions in recent years \cite{fichtner2020rise, lerner2022venture}. In other words, organizations within the network have formed more distinct groups, resulting in a more complex structure. This could be related to the increase in the number of industries in which acquisitions occur, considering that acquisitions mostly happen between companies within the same industry \cite{grullon2019us, schildt2006buys, lee2023identifying}. 

\FloatBarrier

\begin{figure}[ht]
\centering
  \includegraphics[width=15cm,
  keepaspectratio]{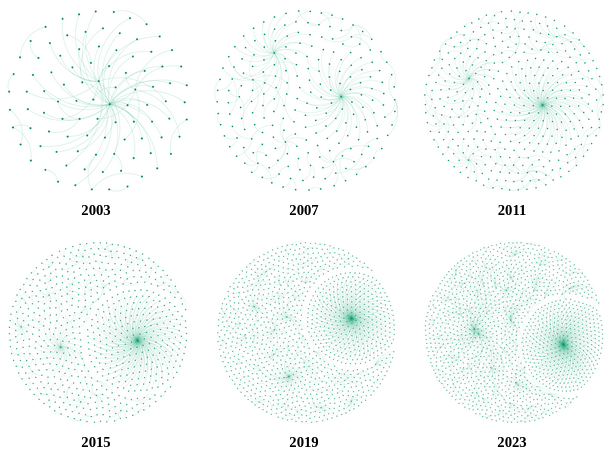}
\caption{Evolution of a sample of the acquisition network over time. This figure further confirms that the network has become sparser and more complex in terms of community structure.}
\label{fig:evolution}
\end{figure}

Figure \ref{fig:evolution} represents the evolution of a sample of the acquisition network over time. We can observe the aforementioned trends in these networks. For instance, the average clustering coefficient of the network has increased, while the density of the network has declined. Moreover, the number of connected components have increased.

\FloatBarrier

\subsection{Cross-city acquisition network}
\begin{figure}[ht]
\centering
  \includegraphics[width=12cm,
  keepaspectratio]{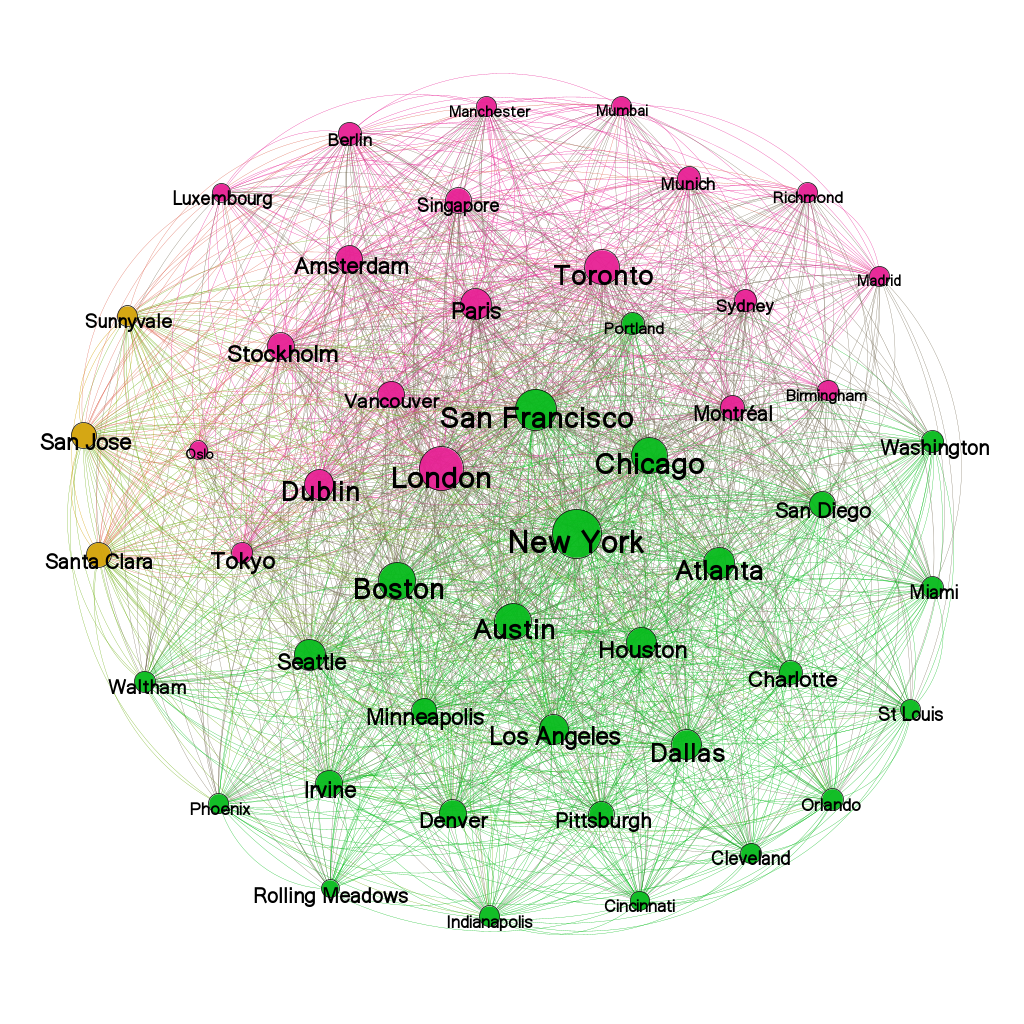}
\caption{A sample of the cross-city acquisition network. The size of each node and its label are proportional to its betweenness and closeness centralities, respectively. Additionally, nodes are colored based on their community detected via the Louvain algorithm. This figure illustrates that the three detected communities mostly consist of cities in the United States, Europe, and California, respectively.}
\label{fig:crossCity}
\end{figure}

Figure \ref{fig:crossCity} displays a sample of the cross-city acquisition network, including 50 cities with the highest numbers of connections (indegree + outdegree) with other cities. In the visualization, node colors indicate their respective Louvain communities, and edge thickness is proportional to their weights. Additionally, node sizes correspond to their betweenness centrality, and the label sizes are proportional to their closeness centrality. According to our analysis, the network reveals three distinct communities: the green community represents cities in the United States, the golden community encompasses cities in California, and the pink community predominantly consists of Canadian and European cities, with a few Asian cities like Tokyo, Singapore, and Mumbai.

\FloatBarrier

\subsection{Centrality analysis}
\begin{figure}[ht]
\centering
  \includegraphics[width=15cm,
  keepaspectratio]{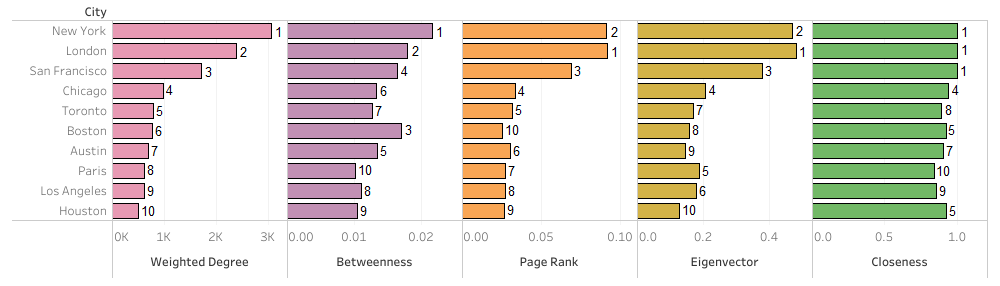}
\caption{Side-by-side bar plot of centrality measures for the top ten central cities within the cross-city acquisition network. The number in front of each bar indicates the ranking of the respective city for each centrality measure. This figure reveals that New York, London, and San Francisco are the most central cities across different centrality metrics.}
\label{fig:centralities}
\end{figure}

Figure \ref{fig:centralities} provides a detailed summary of centrality measures for the top ten central cities in the cross-city acquisition network. In terms of weighted degree centrality, New York exhibits the highest values, aligning with previous studies that assert New York's predominant position in the global economy \cite{currid2006new}. Following New York, London and San Francisco have the second and third highest weighted degrees, respectively. This observation coincides with Budd and Whimster's findings, highlighting London's status as a global financial center \cite{budd1992global}. Additionally, Ester's work suggests that San Francisco, with its accelerators in Silicon Valley, plays a significant role in attracting startups \cite{ester2017accelerators}.

Regarding betweenness centrality, values remain relatively small due to direct connections between most cities. The top three cities' order mirrors that of weighted degree centrality. This ranking indicates these cities' influence in controlling acquisition transactions among other cities.

Figure \ref{fig:centralities} depicts that New York, London, and San Francisco have the highest PageRank values, signifying their active engagement in acquisitions with other influential cities within the network. The top three nodes in eigenvector centrality rankings align with PageRank, except for the first position, which is held by London. This outcome further confirms the strong connections of these cities to central hubs within the network.

In terms of closeness centrality, New York, London, and San Francisco each have a maximum value of 1, signifying direct acquisition transactions with all other cities in the network. Essentially, they have acquired at least one organization from every other city in the network.

\FloatBarrier

\subsection{Cross-border acquisition network}
\begin{figure}[ht]
\centering
  \includegraphics[width=15cm,
  keepaspectratio]{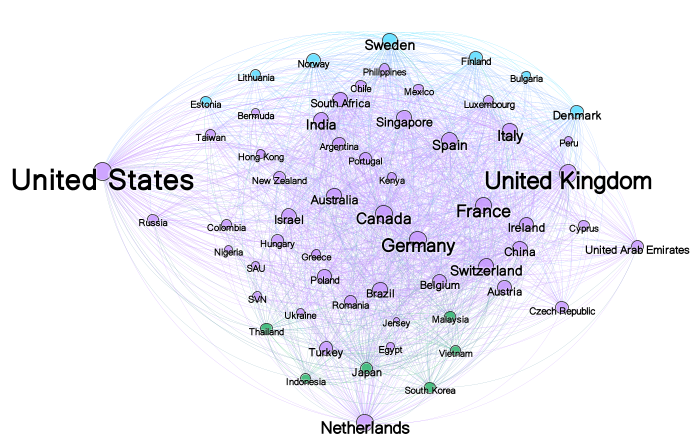}
\caption{A sample of the cross-border acquisition network. The size of each node and its label are proportional to its authority and betweenness centralities, respectively. Additionally, nodes are colored based on their community detected via the Louvain algorithm. This figure highlights the prominence of the United States, United Kingdom, and Germany in global acquisitions.}
\label{fig:crossBorder}
\end{figure}

Figure \ref{fig:crossBorder} showcases a sample of the cross-border acquisition network, featuring the top 60 countries based on their node degrees. Node size reflects their authority centrality, while label size corresponds to their betweenness centrality. Furthermore, nodes are color-coded according to their Louvain community. According to the findings, the United States, United Kingdom, and Germany exhibit the highest authority centralities respectively, indicating their significant influence in the global economy. Additionally, they possess the third-highest betweenness centralities, confirming their pivotal role in facilitating acquisition transactions among other countries. These results align with prior research, reinforcing the dominant position of the United States in the global economy \cite{lew2016america, elson2019key}. Moreover, previous studies have underscored the substantial contributions of the United Kingdom and Germany in cross-border acquisitions \cite{hamill1986foreign, bassen2010m}.

\FloatBarrier

\begin{figure}[ht]
\centering
  \includegraphics[width=15cm,
  keepaspectratio]{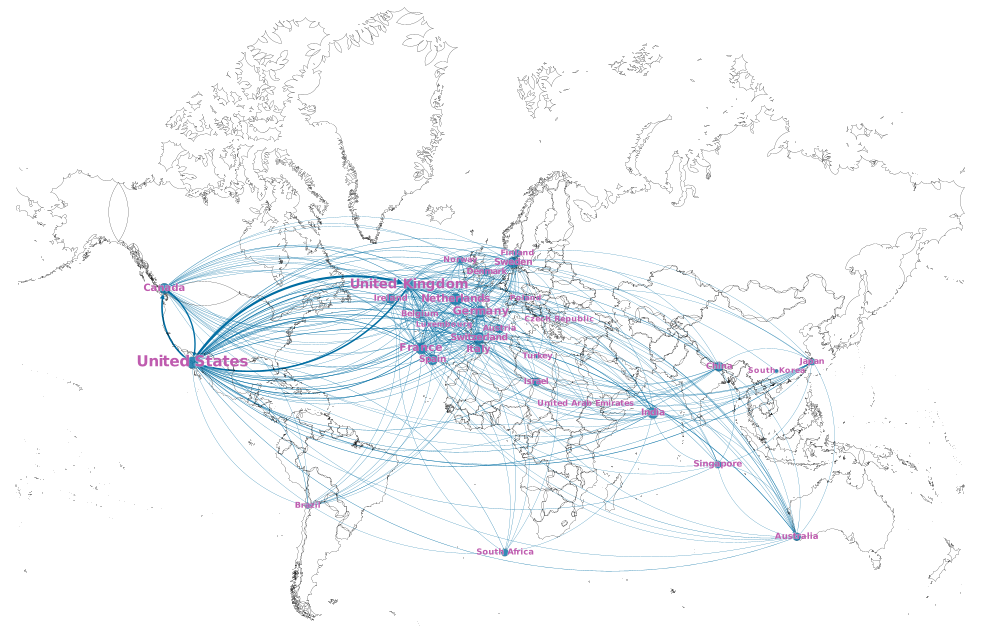}
\caption{A sample of the cross-border acquisition network on the world map. The size of each node and its label are proportional to its degree and betweenness centrality, respectively. The thickness of the edges corresponds to the number of acquisitions between the two countries. This figure further confirms the pivotal roles of the United States and the United Kingdom in the global economy.}
\label{fig:crossBorderMap}
\end{figure}

Figure \ref{fig:crossBorderMap} illustrates the connections between countries in terms of acquisitions on a world map. For visualization purposes, we utilized the undirected version of the cross-border acquisition network involving 30 countries with the highest node degrees. The size of each node corresponds to the number of incoming edges, and the size of its label corresponds to its betweenness centrality. Additionally, the thickness of the edges is proportional to their respective weights.

\FloatBarrier

\section{Discussion}\label{sec5}
Before discussing the implications and limitations of our study and suggesting avenues for future research, we now address the research questions previously mentioned.

\paragraph{RQ1: What features influence the interconnections between organizations within the acquisition network?}
We found that country and category group are the most influential features in the formation of edges (acquisition transactions) between companies. Previous studies also underlined the significant role of geographic proximity and industry on company acquisitions \cite{aybar2009cross, yaghoubi2014acquisition}; however, our findings revealed that the impact of these factors remains significant even when considering a combination of company features. The reason behind this observation could be that acquiring firms are more likely to acquire companies in the countries and industries they have more experience with, particularly their own country and industry \cite{aybar2009cross, galavotti2017experience}. 

\paragraph{RQ2: How has the structure of the acquisition network evolved over time?}
We showed that the acquisitions network has become sparser overall, but the number of clusters within the network has increased, and these clusters have become denser, leading to a more complex structure. These patterns may be due to the absence of monopolies in the global economy and the shift towards industry-based acquisition transactions, which has accompanied the rise in the number of industries \cite{grullon2019us, schildt2006buys, lee2023identifying}.

\paragraph{RQ3: Which cities or countries serve as global economic hubs in terms of acquisitions?}
Our results revealed that New York, London, and San Francisco are the most central cities in global acquisitions, while the United States, United Kingdom, and Germany are the most influential countries in these transactions. This could be due to the fact that these cities and countries have open access policies, advanced infrastructure, and proximity to leading financial markets like Wall Street and Silicon Valley \cite{currid2006new, budd1992global, ester2017accelerators, lew2016america, hamill1986foreign, bassen2010m}.

\subsection{Implications}
The findings of our study provide valuable insights for policymakers aiming to create policies that promote global economic growth and success. Since most acquisition transactions occur within the same industries, policymakers can establish industry-specific regulations to enhance the growth and competitiveness of key sectors. Acknowledging New York, London, and San Francisco as economic hubs, policymakers in other cities can implement strategies modeled after these successful cities to attract acquisitions and investments. Additionally, considering the leading roles of the United States, United Kingdom, and Germany in international acquisitions, policymakers can ease cross-border transactions by lowering regulatory hurdles and encouraging other countries to engage in global acquisitions.

In addition to policymakers, acquiring firms and startups can also benefit from our results when making more informed acquisition decisions. For more successful integrations, acquirers should consider startups from the same country, industry, or age range. These firms can position themselves in economic hubs like New York, London, and San Francisco to gain access to more acquisition opportunities, partners, and investors. They should also consider expanding into key markets such as the United States, United Kingdom, and Germany, which can lead to higher success rates.

\subsection{Limitations}
We acknowledge that our study comes with certain limitations. The Crunchbase data we utilized has its own limitations and biases. There may be delays between the completion of venture capital deals and their disclosure. Some companies may not report the amount of investor capital lost, and startups that fail early might not create a profile on Crunchbase. Additionally, small local businesses and sole proprietorships might be underrepresented since Crunchbase tends to focus on high-growth technology and life sciences companies. Furthermore, as a US-based company, Crunchbase may have a bias towards companies from the US and other Western countries, resulting in less comprehensive data for companies from other regions\footnote{\url{https://news.crunchbase.com/methodology/}}. Therefore, the trends and patterns in acquisitions we observed in this study might be more accurate for Western companies and less generalizable to companies from underrepresented regions.

\subsection{Future work}
Future research can adopt a finer-grained perspective and focus on underrepresented countries to determine whether acquisition patterns in those regions differ from global trends. Another direction for future research is to complement Crunchbase data with local venture capital databases to ensure that local businesses and sole proprietorships are included in the analyses. Additionally, the insights from our study could be utilized in designing models for predicting future acquisitions.

\section{Conclusion}\label{sec6}

In this study, we analyzed global acquisitions with a network approach to gain insights into trends, patterns, and dynamics in this realm. We constructed the acquisition network and its projections on acquirers and acquirees to evaluate interconnections between companies. Our findings revealed that these networks have low densities and comprise a great number of small connected components. By calculating assortativity coefficients, we demonstrated a preference among organizations to acquire companies from the same country, industry, or age bracket as themselves. ERGMs coefficients indicated the significant impact of different company features, such as country, region, city, and category, on the formation of connections between organizations, while founding date did not prove to be the cause of connections. Additionally, we conducted a temporal analysis on the acquisition network, which indicated that the average clustering coefficient has increased over time, while the network has become sparser, comprising more weakly connected components. We computed different centrality metrics for the cross-city acquisition network and found that New York, London, and San Francisco are global economic hubs. Finally, the analysis of the cross-border acquisition network revealed that the United States, United Kingdom, and Germany are the dominant countries in international acquisitions.

\backmatter

\section*{Abbreviations}
  IT, Information Technology; SNA, Social Network Analysis; ERGM, Exponential Random Graph Model; IPO, Initial Public Offering; AUC, Area under the ROC Curve; AI, Artificial Intelligence; CV, Curriculum Vitae; M\&A, Mergers and Acquisitions; NLP, Natural Language Processing; AIC, Akaike Information Criterion.

\section*{Declarations}

\subsection*{Availability of data and materials}

The data that support the findings of this study are available from Crunchbase but restrictions apply to the availability of these data, which were used under the Crunchbase research license for the current study, and so are not publicly available. However, the analysis scripts supporting the conclusions of this paper are available at \url{https://github.com/kalhorghazal/Company-Acquisition-Paper}.

\subsection*{Competing interests}

The authors declare that they have no competing interests.

\subsection*{Funding}

No funding was received for conducting this study.

\subsection*{Authors' contributions}

GK: Data curation, Formal analysis, Methodology, Investigation, Validation, Visualization, Writing- Original draft. BB: Conceptualization, Project administration, Supervision, Writing- Reviewing and Editing. All authors read and approved the final manuscript.

\subsection*{Acknowledgements}

The authors express gratitude to Crunchbase for facilitating access to data.

\bibliography{sn-bibliography}

\end{document}